# TEMPERATURE EFFECTS ON MECHANICAL PROPERTIES OF ZINC DITHIOPHOSPHATE TRIBOFILMS

Karim DEMMOU*, Sandrine BEC, Jean-Luc LOUBET & Jean-Michel MARTIN

Laboratoire de Tribologie & Dynamique des Systèmes - UMR 5513 CNRS/ECL/ENISE
Ecole Centrale de Lyon ; 36 Av. Guy de Collongue - 69134 Ecully Cedex - FRANCE

Abstract:

Nanoindentation tests were performed at several temperatures (24 to 80°C) on one antiwear zinc dialkyl-dithiophosphate (ZDTP) tribofilm using a Nanoindenter XP® entirely set into a climatic chamber. Mechanical properties of the tribofilm were determined using a simple model. AFM observations were conducted to estimate the order of magnitude of the film's thickness. The effect of applied pressure on the elastic properties was demonstrated and taken into account in the present analysis. The use of the $\left(F/S^2\right)$ parameter, independent of contact geometry, revealed a hardness dependency upon temperature. Furthermore, careful AFM observations of the residual pile-up produced by plastic flow around the indents pointed out the evolution of the film deformation process with temperature.

*: Corresponding author.

E-mail address: karim.demmou@ec-lyon.fr

# 1. Introduction

Since the 40's, Zinc dialkyl-dithiophosphates (ZDTP) are high performance lubricant additives which are extensively used because of their good antioxidant and antiwear properties. Many researches have been carried out on these additives in order to better understand how tribofilms are formed and how they act inside contacts. Nevertheless, faced with more and more restrictive rules concerning sulphur and phosphorous contents in lubricants, the aim is now to reduce the use of ZDTP and to develop less pollutant new antiwear additives [1].

Most of these studies deal with the chemical aspect of the formation and action of ZDTP tribofilms [2-5] but few are connected with their own mechanical behaviour [6-9]. The main reason is that mechanical measurements are difficult to perform on such films: ZDTP tribofilms are very thin, with a patchy pattern which results in a great inhomogeneity in shape and thickness. Pads are about 10 to 100 µm long and less than 10 µm wide. Their thickness ranges from a few tens to a few hundreds of nanometers.

Nanoindentation has revealed itself to be an efficient method for measuring near surface mechanical properties and particularly thin films ones. However, in the case of thin films, composite mechanical properties, which take into account both the film's and the substrate's properties, are measured. In order to overcome this difficulty, a simple model was used to extract film's modulus from the composite one [10]. Furthermore, it is known that pressure influences elastic properties of polymer-like materials [11]. Such behaviour has been evidenced for ZDTP tribofilms [6] and has been taken into account in this study.

It is also worth noting that, as far as we know, all published results concern measurements made at ambient temperature although service conditions are around 80°C. In this study, nanoindentation tests were performed on a Nanoindenter XP®, entirely set into a climatic chamber, which permits measurements between -50 and 100°C. So, mechanical properties of ZDTP films have been measured from ambient temperature to 80°C.

Post indentation AFM observations at room temperature have also been performed in order to estimate the thickness of the indented pads and to quantify pile-up around the residual indents.

## 2. Experimental details

### 2.1. Sample preparation

The tribofilm was formed on a Cameron-Plint tribometer using a cylinder (6 mm long and 6 mm in diameter) sliding against a flat, both in AISI 52100 steel. The sliding distance was 7 mm long at a frequency of 7 Hz. The lubricant was a synthetic base oil poly alpha olefin (PAO) with 3% in weight ZDTP. The applied normal load was progressively increased to 350 N during the friction test and then maintained constant (total test duration: 1 hour). The test temperature was kept constant at 80°C.

Optical microscopy observations confirmed the patchy structure of the ZDTP tribofilm.

### 2.2. Nanoindentation tests

Nanoindentation tests were performed with a Nanoindenter XP® using the continuous stiffness measurement method [12-14].

This method consists in superimposing a small displacement oscillation (small enough to generate only an elastic strain) at a given frequency during the indentation test. So, the sample endures small loading-unloading cycles. The simultaneous measurement of the normal force and the contact stiffness allows us to scan mechanical properties, hardness, $H$, and Young's modulus, $E$, continuously along the whole indentation period.

Thanks to the climatic chamber (**Fig. 1 a & b**), nanoindentation tests have been performed at 24, 40, 60 & 80°C. The maximum applied load was 10 mN, at a constant strain rate, $\left(\dfrac{dF}{Fdt}\right)$, of $3.10^{-2}$ s$^{-1}$, where F is the normal load. The optical microscope integrated in the device allowed us to precisely define the indentation location and to verify the position of the residual indents.

The small superimposed displacement had an amplitude of 1.5 nm at a frequency of 32 Hz. The blowing fan frequency of the heating system has been set at 12 Hz during experiments in order to minimize noise.

Corrections have been made to prevent errors due to the geometry of the contact area. Preliminary indents were done on fused silica in order to characterize the tip defect of the Berkovich tip (three side pyramid with an angle of 115,12° between edges). AFM observations were used to obtain topographic information concerning pile-up around the indents, necessary to determine actual contact area.

More than 10 indents have been performed for each temperature at different places on the tribofilm. Some indents were made at different temperatures on the same pads (example on **Fig. 2**).

*2.3. Film's mechanical properties*

The hardness, $H$, is defined as a contact pressure by Eq. 1.

$$H = \frac{F}{A_c} \quad \text{(Eq. 1)}$$

Where $F$ is the normal force applied to the sample, $A_c$ is the actual contact area taking into account pile-up around the indent and tip defect. $A_c$ is defined by Eq. 2 [10].

$$A_c = 26.99 \times \Psi^2 \times (h_{r'} + h_0)^2 \quad \text{(Eq. 2)}$$

where $h_{r'}$ is the contact depth, $h_0$ is the equivalent height of the tip defect and $\Psi^2$ is the ratio between actual contact area (taking into account pile-up) versus theoretical triangular area. The order of magnitude of $\Psi^2$ was estimated using AFM observations. The contact radius, $a$, corresponding to the flat punch which has the same contact area $A_c$ is defined by $A_c = \pi a^2$.

Elastic properties are directly related to the measured normal contact stiffness, $S$, and to the equivalent contact radius, $a$ (Eq. 3 & 4).

$$E^* = \frac{S}{2a} \quad \text{(Eq. 3)}$$

$$\frac{1}{E^*} = \frac{1-v_c^2}{E_c^*} + \frac{1-v_i^2}{E_i^*} \quad \text{(Eq. 4)}$$

where $E^*$ is the apparent reduced modulus, $E_c^*$, $v_c$, $E_i^*$ & $v_i$ are, respectively, the Young's modulus and the Poisson's ratio for sample (c) and indenter (i) $\left(\frac{E_i^*}{1-v_i^2} = 1150 \ GPa\right)$.

In the case of thin films deposited on a substrate, $E_c^*$ is a composite modulus. It is a function of both the film and the substrate elastic properties and of the film thickness.

Several models can be used to extract the film elastic property from the composite one [15-18]. The model developed by S. Bec & al. [10] has the advantage of being simple and it does not need the use of an adjustable parameter nor the knowledge of the Poisson's ratio of the film and the substrate.

S. Bec & al. consider that the film + substrate system can be modelled by two springs connected in series. Considering suitable boundary conditions, the composite modulus can be expressed from Eq. 5:

$$\frac{1}{E_c^*} = \frac{2a}{1+2t/(\pi a)} \left(\frac{t}{\pi a^2 E_f^*} + \frac{1}{2aE_s^*}\right) \quad \text{(Eq. 5)}$$

Where $t$ is the film's thickness, $E_f^*$ and $E_s^*$ are respectively the film's and substrate's reduced Young's moduli. Thus, the knowledge of the two parameters, $t$ and $E_s^*$ (220 GPa for the AISI 52100 steel substrate), allows us to determine the reduced Young's modulus, $E_f^*$, of the film.

*2.4. Anvil effect*

It is known that hydrostatic pressure influences elastic properties of polymers and polymer-like materials [11, 19]. As the film is trapped between the diamond indenter and the steel substrate and cannot flow, the contact pressure i.e. the hardness, has the same order of magnitude as the hydrostatic pressure. This is called Anvil effect: the elastic modulus increases with hydrostatic pressure.

For ZDTP tribofilms, similar effect was observed takes place from a threshold hardness value, $H_0$. Thus, for contact pressure lower than $H_0$, the film's modulus is constant. For contact pressure higher than $H_0$, the film's modulus increases linearly with hardness. This was interpreted like an accommodation process to the severity of contact conditions [6]

This effect was taken into account in the model used to determine the film's reduced Young's modulus, $E_f^*$.

*2.5. AFM observations*

A VEECO® CP-II Atomic Force Microscope was used to perform observations of the indents and particularly of the pile-up at their periphery at room temperature.

Because of the very low load, the residual indents were about 2 or 3 µm in diameter. Even if we could observe them with optical microscopy, AFM permitted high three dimensional resolution and topography measurements. Thereby, pads height could be estimated and pile-up around the indents could be quantified in order to obtain reliable hardness values considering actual contact area.

AFM was used in contact mode with sharp silicon probes coated with gold. For each indent, large images of the whole pad, images of the indent itself and detailed image of the pile-up around the indent were carried out.

3. **Results and discussion**

*3.1. AFM observations*

**Fig. 3** shows an error mode AFM image of two residual indentation prints made on the same pad at 24°C (**Fig. 3a**) and 80°C (**Fig. 3b**). The error observation mode is similar to a derivative image of the topography image. Therefore, changes in colour are linked to changes in surface slope. That is why such images were used to measure actual contact areas and thus to access to $\Psi^2$ value. If pile-up is not taken into consideration, this results in overestimating modulus and hardness values. Because of the irregular "stairs-like" shape of the pile-up around the indents, it was difficult to measure the actual contact area. Nevertheless, taking $\Psi =1.2$, which is the standard value used in similar conditions, gives a good estimation of the contact area. In the following the same value of $\Psi$ was used for all tests, at all temperatures.

On the other hand, topography AFM images were used in order to estimate the order of magnitude of the pad height. For the indents made on the same pad at 24 and 80°C, the height of the pad was estimated around 250±50 nm.

Moreover, it can be notice a reproducible difference between the pile-up formed at 24°C (**Fig. 3a**) and 80°C (**Fig. 3b**). Pile-up, which is stable with time, seems to have been formed by successive pronounced steps at 24°C whereas, at 80°C it seems to result from a more continuous flow of matter. This may denote different deformation processes between these two temperatures. Additional studies are in progress to further investigate this phenomenon (strain rate dependency, larger temperature range, …).

*3.2. Film's modulus*

AFM topography measurements allowed us to estimate the film thickness. This value is one of the parameter necessary to calculate the film reduced modulus, $E_f^*$, from the measured composite modulus with our model. The pads height was estimated at about 250±50 nm. This value was refined by fitting the calculated composite modulus curve with the measured one. The AISI 52100 steel substrate reduced modulus value, $E_s^*$, was taken to 220 GPa .

**Fig. 4** shows the evolution of the film modulus and hardness versus plastic depth for one indent at 24°C. The film modulus and the hardness are first constant at respective values $E_{f_0}^*$ and $H_0$, until a depth $h_{r'_0}$ and then increase. Similar curves were obtained for all indents at all temperatures.

For hardness values higher than the threshold value $H_0$, the film's modulus curve can be fitted, as expected from [6], by a linear function of the hardness, $H$ (Eq. 6)

$$E_f^* = \frac{E_{f_0}^*}{H_0} H \qquad \text{Eq. 6}$$

As an illustration, **Fig. 5** shows the evolution of the film modulus versus hardness for two indents performed on the same pad at 24°C and 80°C. For all tests the film modulus, $E_{f_0}^*$, was found to be independent of temperature, at a value $E_{f_0}^* \approx 39 \pm 4$ GPa. On the opposite, the threshold contact pressure, $H_0$ was found to be temperature dependent. The threshold value of contact pressure at 24°C is about twice the threshold value at 80°C.

**Fig. 6a & 6b** show the evolution of the composite modulus, $E_c^*$, versus the indentation depth, $h_{r'}$ calculated from using $E_{f_0}^*$ value in our model. From the black curve (symbol ×) on **Fig. 6**, an increase of the composite Young's modulus is observed as the contact depth increases. For contact pressure under the threshold value, this increase is only due to the influence of the substrate. That is why, for shallow indentation depth, the composite modulus calculated with our model and a constant film modulus $E_{f_0}^*$ fits well the measured one (red curves, symbol ○).

For deeper indentation, the calculated curve deviates from the measured composite modulus. The effect of hydrostatic pressure becomes significant.

From the threshold pressure $H_0$, which corresponds to the indentation depth $h_{r'_0}$ where the model deviates from the experimental data, the film's modulus is considered to increase with the pressure as defined in Eq. 6. Using this increasing value for the film modulus gives a better fit of the calculated composite modulus curve with the measured one for indentation depth deeper than $h_{r_0}$, (blue curves, symbol △).

*3.3. Temperature effects*

In order to better evidence the temperature effects on mechanical properties, the $\left(F/S^2\right)$ parameter [normal load divided by contact stiffness squared] versus stiffness was chosen. It was plotted versus the contact stiffness, $S$, which is proportional to indentation depth, because it is a direct measurement not affected by contact geometry considerations. Thus, $S$ is more accurate than indentation depth for measurements made on the same pads.

$\left(F/S^2\right)$ parameter is independent of geometrical considerations inherent to nanoindentation contact modelling [20]. Moreover, $\left(F/S^2\right)$ parameter is proportional to $\left(H/E^2\right)$ ratio. As it was just shown that the Young's modulus of the ZDTP film is constant in the studied temperature range, variations in this parameter are due to variations in hardness, $H$.

The influence of temperature on $\left(F/S^2\right)$ parameter versus contact stiffness, S, is shown on **Fig. 6 & 7**.

From **Fig. 7**, it can be observed, for two indentations made on the same pad at 24°C and 80°C, that the $\left(F/S^2\right)$ parameter decreases rapidly for small stiffness values (corresponding to shallow indentation depths). For high stiffness values, it tends to an asymptotic value corresponding to a greater influence of the substrate on stiffness. For low stiffness, the $\left(F/S^2\right)$ parameter measured at 24°C is significantly higher than at 80°C. As the film modulus is the same at both temperatures, this corresponds to a significant difference in hardness values: ZDTP film's hardness is about twice lower at 80°C than at 24°C.

**Fig. 8** shows $\left(F/S^2\right)$ parameter for several indentations made at different temperatures on different pads. The decrease in hardness with increasing temperature is confirmed independently of the chosen pad. Scattering of values might be due to thickness difference between pads. It is more pronounced at high temperature where experiments are more delicate to conduct.

## 4. Conclusions and prospects

First, nanoindentation tests were made on a patchy ZDTP tribofilm at several temperatures, up to 80°C. The highest temperature, 80°C, is near service temperature. Careful analysis of the obtained data allowed us to extract the film's Young's modulus. Its independence with temperature variations in the studied range which extends from 24 to 80°C was shown. On the other hand, the study of the $\left(F/S^2\right)$ parameter versus stiffness allowed us to point out an important decrease in hardness while the temperature increases.

Second, AFM observations have been used to quantify the pile-up around indentation print and to estimate the height of the pad. These parameters were necessary to extract the Young's modulus of the film from composite properties. Moreover, the observation of the pile-up around the indents revealed a difference between deformation processes depending on temperature.

Thus, the next step of this study will be the investigation of the viscous-plastic properties of these films against temperature in order to better understand the different behaviours highlighted by AFM observations and to try to relate them to physical changes due to temperature increase.

**Figure Captions**

**Fig. 1:** Nanoindenter XP®
  *a)* Climatic chamber. On the right side stands the heating-cooling system. This system can reach temperatures from -50°c to 100°C. Liquid Nitrogen is used to cool the chamber.
  *b)* Integrated in its climatic chamber, it stands on a pneumatic table in order to avoid vibrations. During experiments air flow is made as small as possible and is blown at low velocity in order to minimize perturbations near the device. Blowing fan frequency is tuneable.

**Fig. 2:** Optical observation of two indents on same pad at 24 & 80°C.
Color homogeneity corresponds to thickness homogeneity.

**Fig. 3**: AFM error mode images of two indents on the same pad.
  a)  indent made at 24°C (4x4µm)
  b)  indent made at 80°C (3x3µm).

**Fig. 4:** Representative curves of Young's modulus and hardness of the film versus plastic depth. Film modulus and hardness are first constant then increase from a threshold value of plastic depth $h_{r'_0}$.

**Fig.5:** Film modulus versus hardness for two indents performed at 24°C and 80°C on the same pad

**Fig. 6:** Composite modulus taking into account film and substrate ones as a function of indentation depth,
a) at 24°C & b) at 80°C. Indentations were performed on the same pad.
  ✕ – Experimental composite modulus
  ○ – Composite modulus calculated with Bec's model
  ◇ – Composite modulus calculated with Bec's model and taking into account a film's modulus growing linearly with applied pressure from a threshold value

**Fig. 7:** Load divided by stiffness squared versus stiffness for two indents on the same pad. For shallow indentation depth, the F/S² parameter is twice lower at 80°C than at 24°c.

**Fig. 8:** Load divided by stiffness squared versus stiffness for several indentation made at different temperatures (24 (Black), 40 (Blue) 60 (Brown) & 80°C (Red)) on different pads.

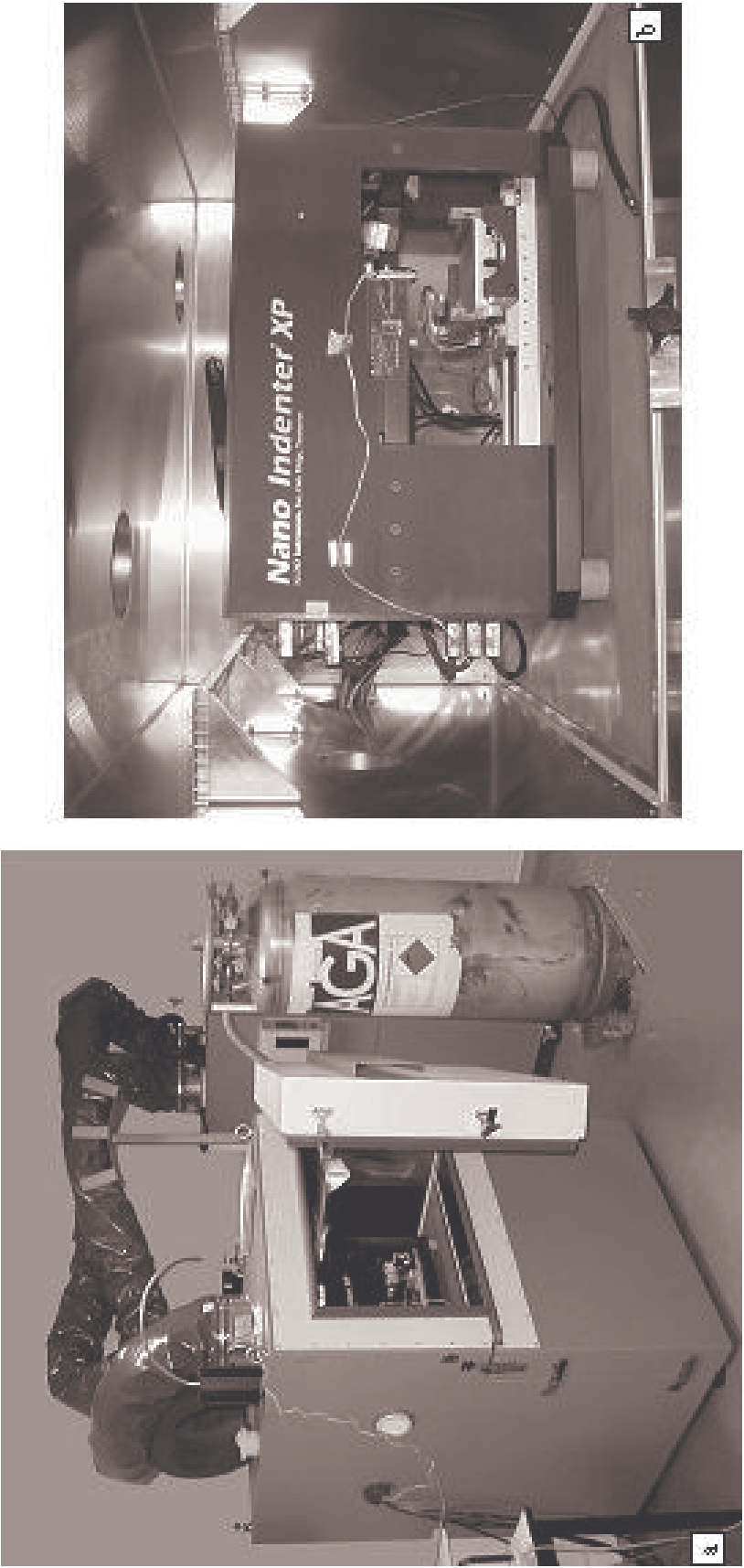

**Fig. 1**

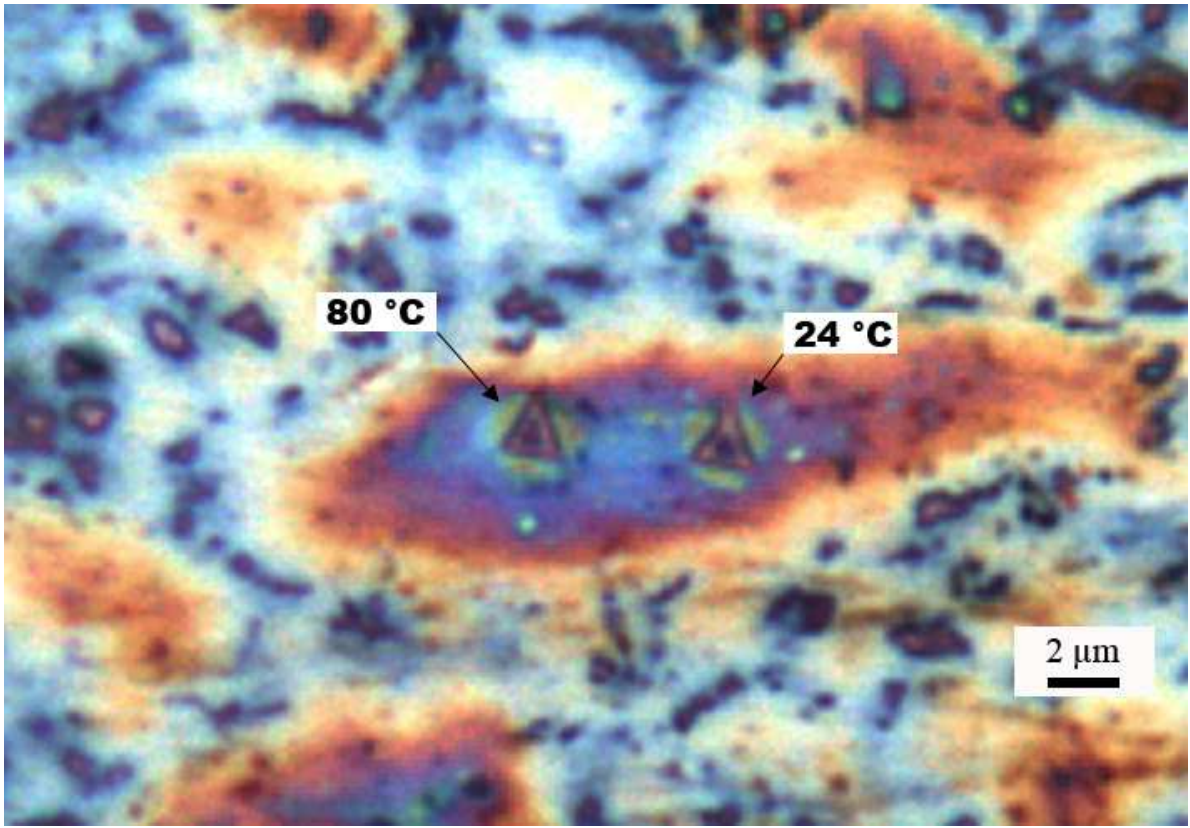

**Fig. 2**

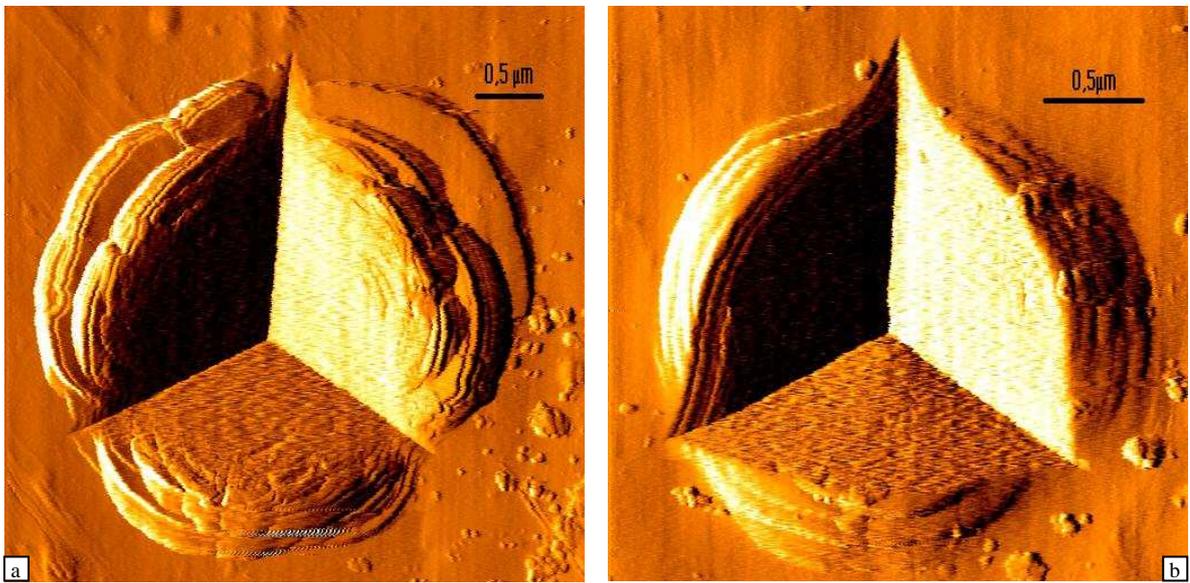

**Fig. 3**

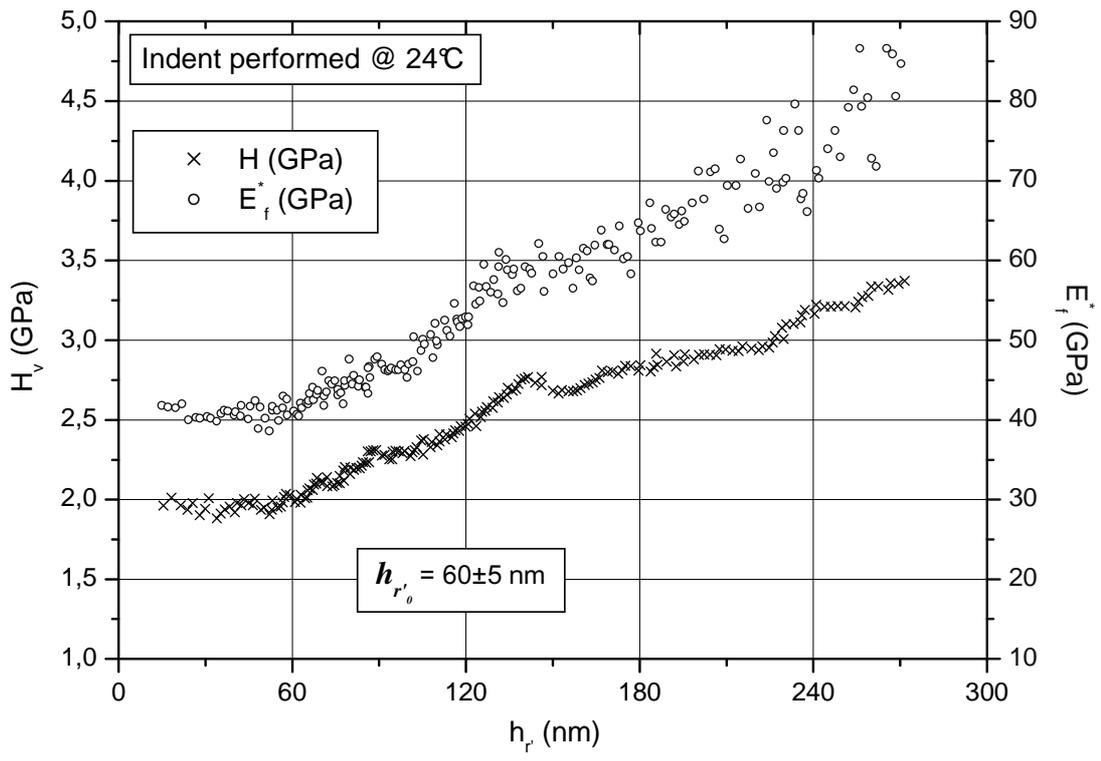

**Fig. 4**

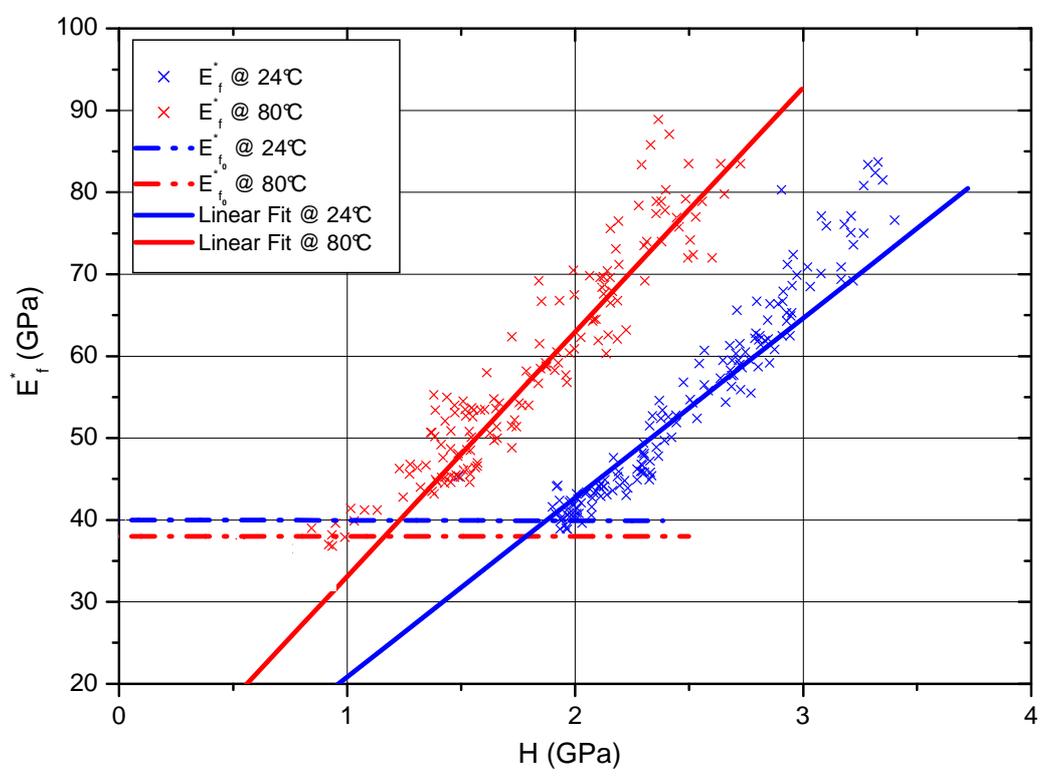

**Fig. 5**

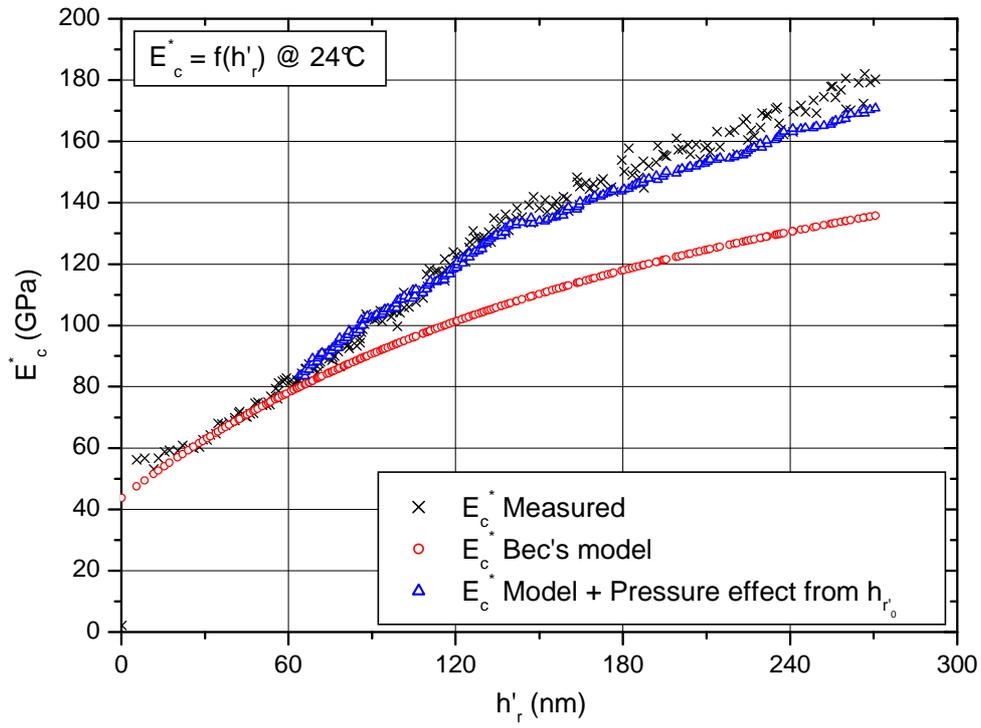

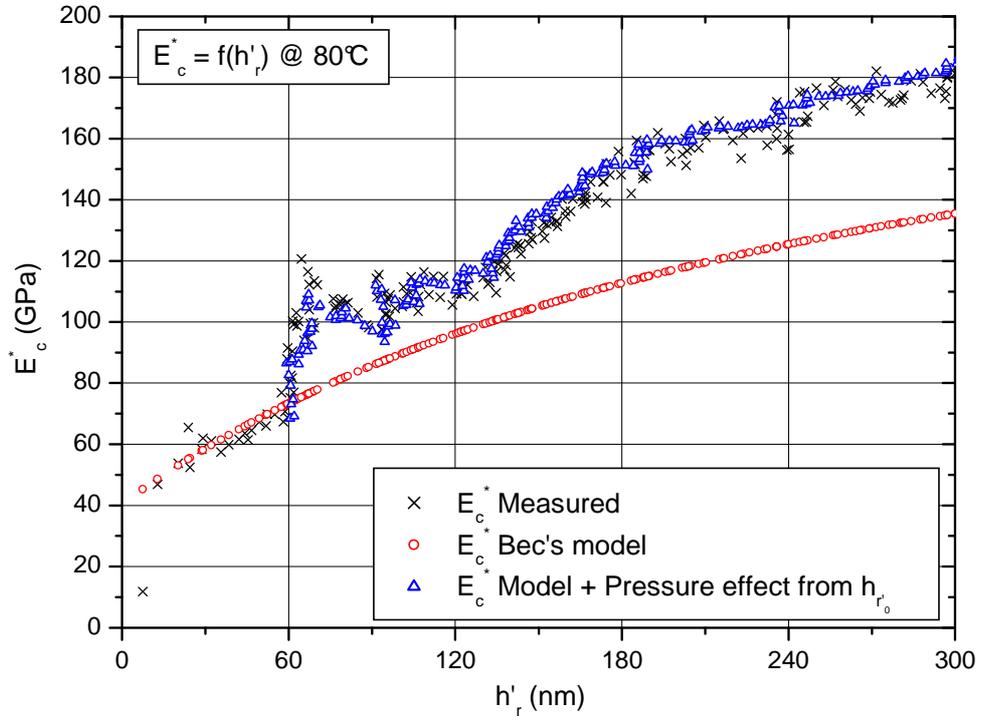

**Fig. 6**

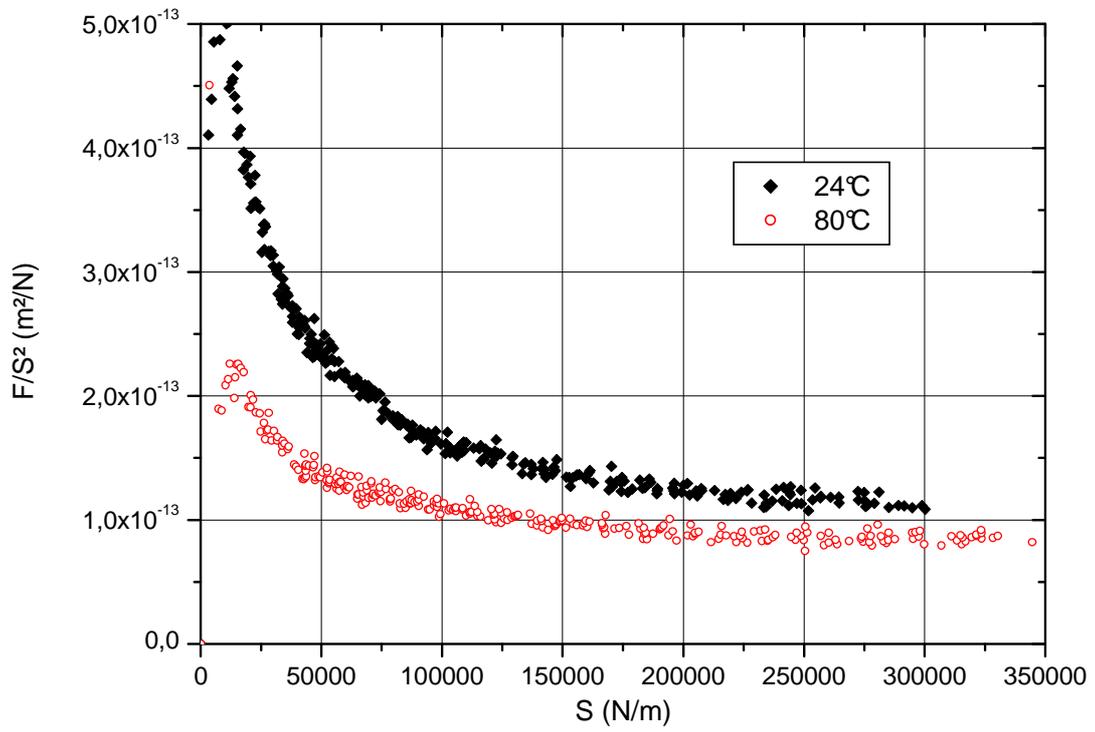

**Fig. 7**

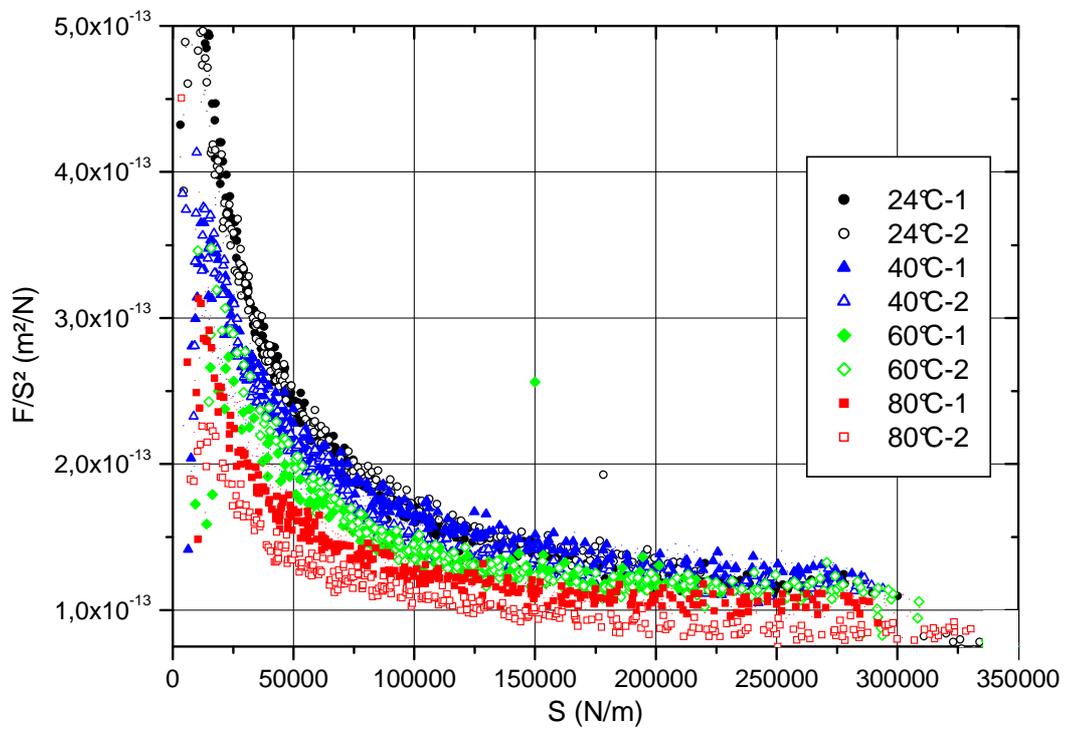

**Fig. 8**